# UofA-Truth at Factify 2022 : Transformer And Transfer Learning Based Multi-Modal Fact-Checking

Abhishek Dhankar[1], Osmar R. Zaïane[1] and Francois Bolduc[1]

[1]*University of Alberta, Edmonton, Alberta, T6G 2E8, Canada*

**Abstract**
Identifying fake news is a very difficult task, especially when considering the multiple modes of conveying information through text, image, video and/or audio. We attempted to tackle the problem of automated misinformation/disinformation detection in multi-modal news sources (including text and images) through our simple, yet effective, approach in the FACTIFY shared task at De-Factify@AAAI2022. Our model produced an F1-weighted score of 74.807%, which was the fourth best out of all the submissions. In this paper we will explain our approach to undertake the shared task.

**Keywords**
fake news, multi-modal, De-Factify@AAAI2022, FACTIFY shared task

## 1. Introduction

Humankind has dealt with misinformation since time immemorial [1]. However, never in human history have people had access to the amount of information that they have today. The Internet is the primary reason for the easy access to this information. It has given people the ability to access information from all over the world and from innumerable sources. However, this deluge of information has brought with it the problem of misinformation/disinformation/fake news. Never before have we had more efficacious means to disseminate deceptive fallacies, falsehood that is unfortunately believed and is wrongfully, and sometimes dangerously impacting people.

While there are many definitions of Fake News, for the purposes of this paper Fake News can be defined as a news piece, social media post, etc., which contains claim(s) that can be refuted by information put out by "reputable organizations". Such organizations may include, but are not limited to, government bodies, news outlets which score high on Media Bias/Fact Check's Factual Reporting scale [2] or professional fact-checking organizations which are verified signatories of the International Fact-Checking Network (IFCN) code of conduct [3] [4]. This definition of fake news off-loads the responsibility of determining what exactly fake news is, on to expert fact-checkers or domain experts, and allows Artificial Intelligence (AI) to deal with the more manageable problem of determining whether claim(s) made in a news piece is entailed, not entailed or refuted by a corresponding news piece from a reputable source.

Fake news can cause real world harm as is being seen during the COVID-19 pandemic: misinformation has led to vaccine hesitancy, which is directly tied to increased chances of mortality due to COVID-19 [5]. Fact-checking or determining whether a news piece contains fake claims is the first step in countering such fake news. Furthermore, it is not only important to detect and counteract fake news, but to do so in a timely manner. Given the large amount of information generated on social media sites every-day and the time constraints that online fact-checking operates under, it is imperative that automated methods of misinformation detection are developed to aid in the manual fact-checking of fake news.

In general, the information generated and distributed on the Internet is multi-modal, i.e., consisting of text, images, audio-visual, etc. Often times information is conveyed via a combination of two or more modes, for instance, memes, pieces of information rapidly spread among users, are often a combination of text and image/short video, where text is overlaid on the image or short video (also called a gif). Thus, an automated method should be able to take advantage of all the modes of information available to fact-check a claim.

The shared task FACTIFY, in conjunction with the AAAI conference, attempts to aid in the development of automated multi-modal fact-checking by introducing a dataset which consists of multi-modal claims and corresponding supplementary information or documents, using which said claims need to be fact-checked [6]. Each multi-modal "claim" consists of a short sentence or phrase and an associated image, which may or may not have overlaid text. The corresponding supplementary information or "document" consists of a sentence or sentences from a reputable source with an accompanying image. The task is to determine whether the claim text and claim image are individually entailed, not entailed, or refuted by the corresponding document text and image pair. Depending on that, and defined by the shared task organizers, there are five possible labels which each claim text and image pair can have:

- Support_Multimodal: Both claim text and image are entailed
- Support_text: Text is entailed, but the image is not
- Insufficient_Multimodal: Claim text is not entailed, but claim image is
- Insufficient_Text: Neither claim text nor claim image is entailed
- Refute: Both claim text and image are refuted

Our team "UofA-Truth" participated in the shared task and secured the $4_h$ position with a weighted F1-score of 74.807%, just ≈ 2 F1 points behind the top submission. In this paper we shall describe our simple yet effective automated fact-checking model.

## 2. Related Works

The dataset used to train and test our model was released under the shared task FACTIFY, which is a part of the workshop De-Factify at the AAAI 2022 conference [6]. The dataset consists of a total of 50, 000 claim and document pairs, which are divided into train, validation and test sets of sizes 35, 000(70%), 7, 500(15%) and 7, 500(15%) respectively.

The entailment aspect of the shared task is similar to "Stance Detection", which can be defined as the classification of the stance of the producer of a news piece with respect to an unverified claim [7]. In the context of the shared task, the unverified claim is the claim text and image pair, and the news piece is the document text and image pair.

Stance Detection is an important part of Fake News detection and was notably used in the Fake News Challenge - 1 (FNC-1) [8]. This challenge was similar to the FACTIFY shared task, except FNC-1 only dealt with text entailment or stance detection, unlike FACTIFY which deals with multi-modal entailment. FNC-1 introduced a dataset which consisted of a headline and a body of text, which may be from the same article or different articles. Depending on the stance of the body of text with respect to the headline, the text-headline pairs were to be classified into any of the following classes:

- Agrees: The body of text agrees with the claim(s) made in the headline
- Disagrees: The body of text disagrees with the claim(s) made in the headline
- Discusses: The body of text and headline are referring to the same subject, but the body does not take any stance or position on the claim(s) made in the headline
- Unrelated: The body of text is not related to the claim(s) being made in the headline

FACTIFY's not-entail class can be considered similar to a combination of Unrelated and Discusses classes of FNC-1, while entails and refutes classes can be considered similar to FNC-1's Agrees and Disagrees classes respectively.

This similarity between the two tasks led us to draw inspiration from the UCL Machine Reading team's submission to the FNC-1's challenge, which performed 3 best among the 50 submissions to the challenge [9]. In their submission the UCL team, Riedel et al., describe their approach as a "simple but tough-to-beat baseline" for stance detection. As explained above, there are two inputs for this task - a headline and a body of text. Riedel et al. calculated the Term Frequency (TF) vectors and Term Frequency - Inverse Document Frequency (TF-IDF) for both the headline and the body of text on the basis of the 5, 000 most frequent words. The TF vectors of the two inputs are concatenated with result of the cosine similarity of between the TF-IDF vectors of the headline and body, as shown in Figure 1. The resultant vector of length 10, 001 is then fed as input into a shallow Multi Layer Perceptron (MLP) network, which has a softmax output of length four, one for each class in the FNC-1 task.

Given the similarity of the tasks being solved in FNC-1 and FACTIFY, we adopted the manner of concatenation of the cosine similarity and vector representations of the header and body as explained in [9]. Instead of using TF for vector representations and TF-IDF for cosine similarity calculation, we used Sentence-BERT in lieu of both TF and TF-IDF vectors to determine entailment between the claim text and document text [10]. Sentence BERT is a transformer based model which has been specifically trained to represent sentences and paragraphs and therefore may be better for sentence representation than other methods which usually compute the average across vector representations of individual words to obtain sentence or paragraph representations. To determine entailment between claim and document images, we used a pre-trained instantiation of the Xception architecture [11] available in Keras [12], which had been trained on JFT-300M dataset [13].

The FNC-1 challenge had two other submissions which performed better than [9], however, we concluded that those were more complicated architectures and might hamper the scalability and time complexity of our model. For instance, Pan et al., who submitted the winning model [14], had an ensemble model which consisted of a deep learning model and a tree based ensemble model as implemented in Xgboost [15]. The outputs of the two models were weighted equally to produce the final predictions. Hanselowski et al. also implemented an ensemble model which consisted of five Neural Network models. The final prediction was made through majority voting. Despite the increased complexity, Team UCL Machine Reading's model performance was within ≈ 1 point of the top two submissions.

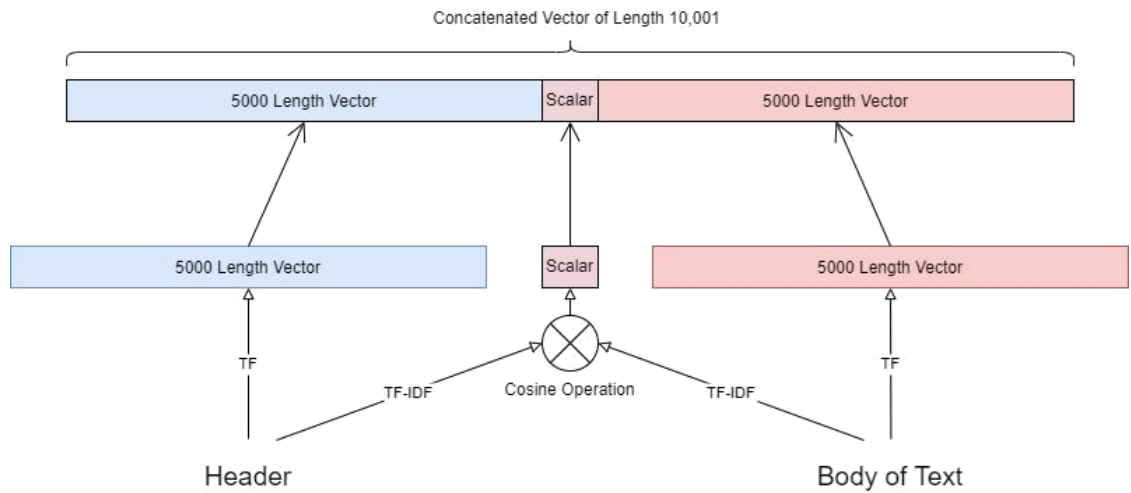

**Figure 1:** Concatenated Vector Representation. Adapted from "A simple but tough-to-beat baseline for the Fake News Challenge stance detection task", by Riedel et al.

## 3. Methodology

The FACTIFY shared task's classes (Support_Multimodal, Support_text, Insufficient_Multimodal, Insufficient_Text, Refute) are composed of a combination of text and image entailment classes. For instance, if text entailment, non-entailment and refutation are represented by _0, _1, _2, and image entailment, non-entailment and refutation are represented by ℐ_0, ℐ_1, ℐ_2 respectively; then the shared task's classes can be reformulated as as a combination of text and image entailment labels as shown in Table 1. It is important to note here that all combinations of text entailment labels and image entailment labels are not present in Table 1. For instance, combinations such as _0 & ℐ_2 do not exist. The lacking combinations are treated when consolidating the labels after classification. This is explained in Section 3.4.

It can be clearly seen that the shared task can now be broken down into two sub-tasks; namely, text entailment and image entailment, where text entailment consists of classes _0, _1 and _2, and image entailment consists of classes ℐ_0, ℐ_1, ℐ_2. These new classes are the combination of the original class labels as shown in Table 2 and Table 3 for the text entailment and image entailment tasks respectively. Once the dataset is rearranged according to the sub-task labels, we end up with one dataset for each sub-task.

We now define Text Entailment as a task of predicting the document text's stance towards the

**Table 1**

FACTIFY Task Labels & Corresponding Text and Image Entailment Labels

| FACTIFY Task Label | Text Entailment Label | Image Entailment Label |
|---|---|---|
| Support_Multimodal | $\_0$ | $\mathcal{I}\_0$ |
| Support_text | $\_0$ | $\mathcal{I}\_1$ |
| Insufficient_Multimodal | $\_1$ | $\mathcal{I}\_0$ |
| Insufficient_Text | $\_1$ | $\mathcal{I}\_1$ |
| Refute | $\_2$ | $\mathcal{I}\_2$ |

**Table 2**

Text Entailment Task Labels in Terms of Original FACTIFY Task Labels

| Text Entailment Label | FACTIFY Labels |
|---|---|
| $\_0$ | Support_Multimodal & Support_text |
| $\_1$ | Insufficient_Multimodal & Insufficient_Text |
| $\_2$ | Refute |

**Table 3**

Image Entailment Task Labels in Terms of Original FACTIFY Task Labels

| Image Entailment Label | FACTIFY Labels |
|---|---|
| $\mathcal{I}\_0$ | Support_Multimodal & Insufficient_Multimodal |
| $\mathcal{I}\_1$ | Support_text & Insufficient_Text |
| $\mathcal{I}\_2$ | Refute |

claim text, and Image Entailment as a task of prediction the document image's stance towards the claim image.

### 3.1. Preprocessing

Image preprocessing is done by resizing all the images to (256, 256, 3) size with bilinear interpo-lation as implemented in image_dataset_from_directory in Keras [12]. Thereafter, all the pixel values are scaled to a a range of 0 to 1.

Text preprocessing involves removing urls from all claim and document texts with the help of the Preprocessor library [17].

### 3.2. Vector Representations

The preprocessed inputs (text and images) need to be converted into vector representations so that they can be presented as input for a classifier.

The preprocessed images are converted into vectors of size 2048 each, by using the pre-trained Xception model in Keras [12]. This can be achieved by setting include_top attribute to False and pooling attribute to 'avg'. Setting include_top to False removes the fully connected layer at the end of the model and exposes the output of the second to last layer. Setting pooling to 'avg' ensures that a global pooling average is applied to the 3D output of the second last layer of Xception, to convert it into a 1D output. Since the Xception model has been trained on a massive dataset for a general image classification task, it can be reasonably assumed that the output of the second to last layer captures information which may be useful for downstream tasks such as image entailment.

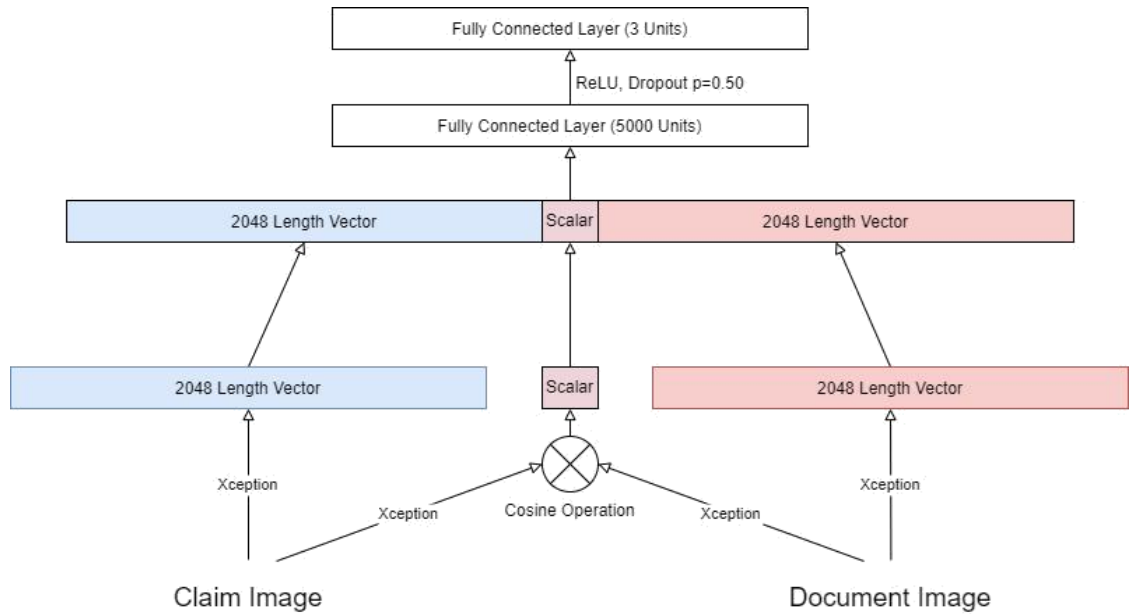

**Figure 2:** Image Entailment Classifier Architecture

The preprocessed texts are converted into a vectors of length 384 each by using the pre-trained Sentence-BERT model [10].

The cosine similarity of the vector representations of claim and corresponding document images is calculated and concatenated in the manner shown in Figure 2. This creates a concate-nated representation for each claim and corresponding document image of size 4097. Similarly, the concatenated representation of claim and corresponding document text of size 769 is created through the same procedure of cosine similarity calculation and subsequent concatenation as shown in Figure 3.

### 3.3. Classifiers

The vector representations are now used for training the classifiers for the image and text entailment tasks. Different classifiers are used for the image and text entailment tasks.

As shown in Figure 2 the image entailment classifier consists of a single fully connected hidden MLP layer of 5000 units, ReLU activation with a dropout probability of 0.5. The output of this layer then feeds into a fully connected output layer of 3 units, one for each class label (entailment, non-entailment and refute), and a sigmoid activation function.

On the other hand, as shown in Figure 3 the text entailment classifier consists of two fully connected layers of 450 units each, ReLU activation functions, 2 activity regularizers, and a dropout probability of 0.55 for the first layer and 0.4 for the second layer. The output of the two hidden layers then feeds into the fully connected output layer 3 units with sigmoid activation.

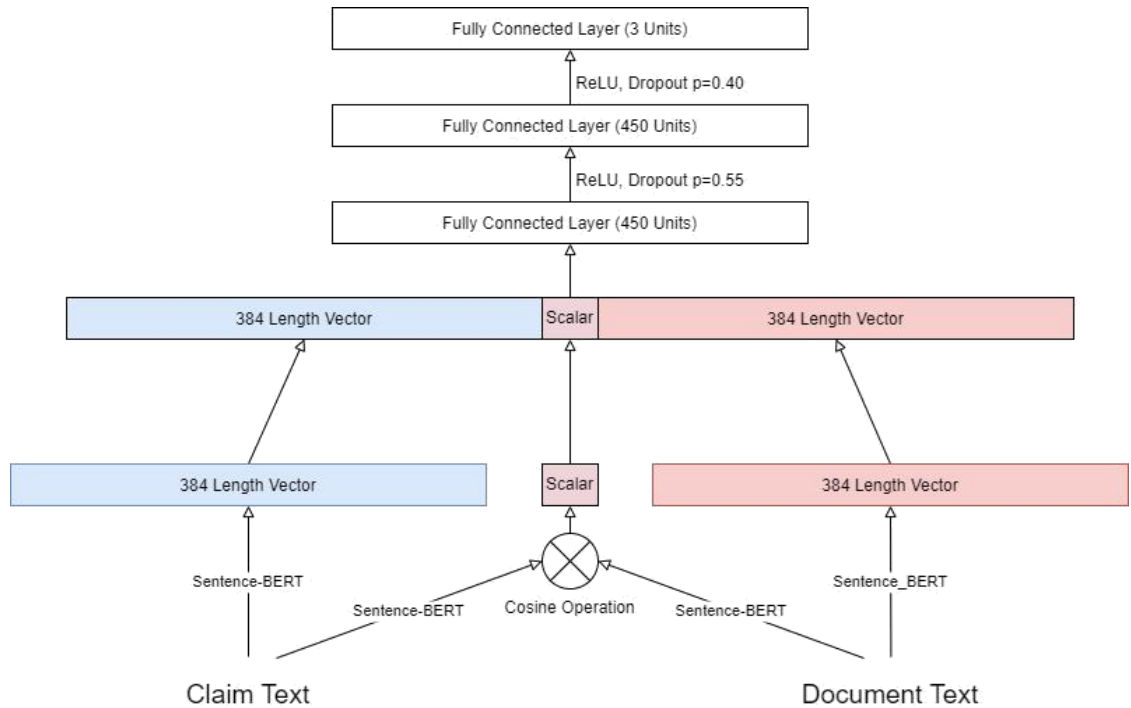

**Figure 3:** Text Entailment Classifier Architecture

The Cross Entropy loss is calculated after performing the softmax operation on the outputs of both the classifiers.

### 3.4. Label Consolidation

The output of the image classifier classifies every pair of claims and document image into one of the three labels $\mathcal{I}\_0$, $\mathcal{I}\_1$ and $\mathcal{I}\_2$. Similarly, for claim and document text pairs.

The pairs of image and text entailment labels belonging to the same data-point are combined and then converted into the original FACTIFY task labels (namely, Support_Multimodal, Support_text, Insufficient_Multimodal, Insufficient_Text, Refute) according to Table 1.

However, it is possible that the combination procedure may produce pairs of entailment labels which do not have any corresponding FACTIFY task label. For instance, ($\_0$, $\mathcal{I}\_2$), ($\_1$, $\mathcal{I}\_2$), ($\_2$, $\mathcal{I}\_0$), ($\_2$, $\mathcal{I}\_1$), are four such invalid pairs of labels. We thus have to change such label pairs into valid label pairs. We do so by using a heuristic as described in Table 4(A). If one of claim text or claim image is entailed, i.e., $\_0$ or $\mathcal{I}\_0$, it is unlikely that the other claim mode will be refuted by the document, hence, the latter's label needs to be changed to not-entailed, i.e., $\_1$ or $\mathcal{I}\_1$. If however, one of the claim text or image is refuted by the corresponding document then it is unlikely that the other claim mode will have uncertain entailment, hence the latter's label should be converted to refuted as well, i.e., $\_2$ or $\mathcal{I}\_2$ . Thereafter, we can calculate the final weighted F1 accuracy on the Test set.

**Table 4**
Heuristics for invalid label conversion

| Invalid Label Pair | Valid Label Pair | Invalid Label Pair | Valid Label Pair |
| --- | --- | --- | --- |
| ( $\_0, \mathcal{I}\_2$) | ( $\_0, \mathcal{I}\_1$) | ( $\_0, \mathcal{I}\_2$) | ( $\_0, \mathcal{I}\_0$) |
| ( $\_1, \mathcal{I}\_2$) | ( $\_2, \mathcal{I}\_2$) | ( $\_1, \mathcal{I}\_2$) | ( $\_1, \mathcal{I}\_1$) |
| ( $\_2, \mathcal{I}\_0$) | ( $\_2, \mathcal{I}\_2$) | ( $\_2, \mathcal{I}\_0$) | ( $\_2, \mathcal{I}\_2$) |
| ( $\_2, \mathcal{I}\_1$) | ( $\_1, \mathcal{I}\_0$) | ( $\_2, \mathcal{I}\_1$) | ( $\_2, \mathcal{I}\_2$) |

**A:** Invalid to Valid Label Pair Conversion **B:** New Invalid to Valid Label Pair Conversion

## 4. Results & Discussion

Our team, UofA-Truth, secured the $4^h$ position on the leaderboard, with an F1-score of 74.807% on the final evaluation. However, the confusion matrix, shown in Figure 4, reveals more fine grained details about our model's performance on the test set.

The model performed worst on the Insufficient_Multimodal category. It can be clearly seen from the matrix that a large number (304) of data-points with ground truth Insuffi-cient_Multimodal were incorrectly classified as Support_Multimodal. Since the only difference between the two classes is text entailment, it is possible that the model is unable to differentiate between text entailment and non-entailment. This may be because the claim and document texts might have common words or might even talk about tangential or similar topics, but do not reach the threshold of text entailment.

Furthermore, the model did not perform well on the Support_text class. A significant number (202) of data-points belonging to Support_Text were mis-classified as Support_Multimodal. Again, given that the only difference between the two classes is in image entailment, it follows that a model would find it hard to differentiate between the two. Similarly, for the Support_Text and Insufficient_Text.

The model performs best on the Refute class despite the fact that the said class had the fewest data-points in the training set. Very few of its data-points are mis-classified as other classes, and vice-versa. This may be because a significant number of the data-points belonging to the Refute class have been take from fact-checking websites. This fact may set such data-points apart from other claim-document pairs. For instance, document images and corresponding claim images of the Refute class tend to be identical because fact-checking websites almost always provide a screenshot of the fake news/social media posts they debunk in their articles. They may even overlay images of news pieces they fact-check with a digital stamp, indicating their logo or whether the news piece was true or fake. They usually clearly state the gist of the fake news they debunk, at the beginning of every article, often times quoting said fake news verbatim. Such peculiarities may make data-points belonging to the Refute class easy to discern.

Heuristics mentioned in Section 3.4 can be changed to improve the weighted F1-score on the test set. It is possible that the image entailment model merely learns to determine similarity between claim and document image pairs. Thus, it may be better to have a heuristic which changes the invalid label pairs into valid label pairs by changing the image entailment label to be the same as the text entailment label. Therefore, after the competition results we changed the heuristic for invalid label pair to valid label pair conversion as per the new heuristics shown in Table 4(B). These modified heuristics improve the final F1-score from 74.807% to 75.183%. As can be seen by the confusion matrix in Figure 5, the new heuristic reduces the classification accuracies of the Insufficient_Multimodal and Support_Text classes, for the benefit of the other classes. Other than that, the overall dynamics remain the same as in Figure 4. It could be possible to continue adjusting these heuristics to obtain even better results but have not experimented further.

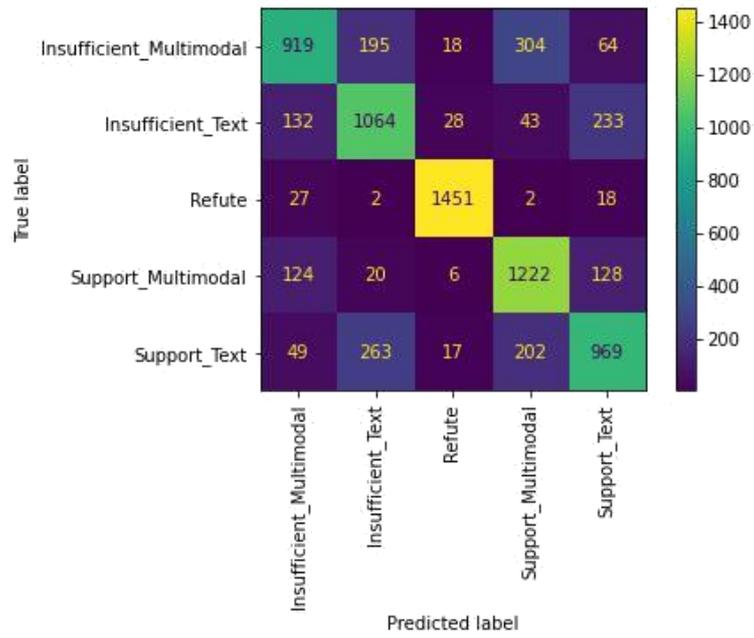

**Figure 4:** Confusion Matrix for Original Heuristic

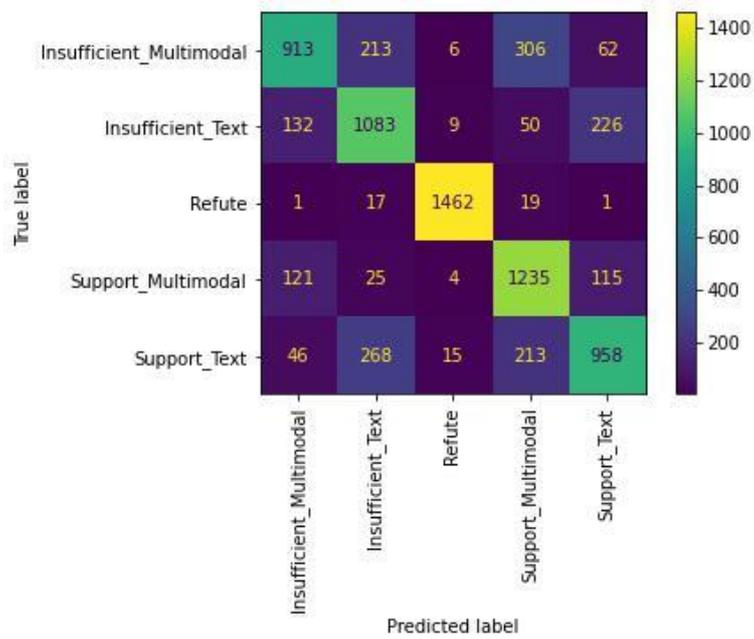

**Figure 5:** Confusion Matrix for Modified Heuristic

## 5. Conclusion

In this paper, we introduced a simple, yet effective method of multi-modal fake news detection. We divided the main task into two sub-tasks; namely, text entailment and image entailment. Thereafter, we used pre-trained Xception network and Sentence-BERT to get vector representations of images and text respectively. We then used these vector representations for classifications tasks of image and text entailment by adapting the approach introduced by Riedel et al. in their submission to the FNC-1 task. Finally, we consolidated the prediction of the two sub-tasks of image and text entailment to get the final predictions. We used the model thus created to make predictions on the test set, and our team's submission achieved the $4_h$ position on the leader board with a 74.807% weighted F1-score.

## Acknowledgments

This work was partially funded by a Collaborative Health Research Project grant from the Canadian Institutes of Health Research (CIHR) and the Natural Sciences and Engineering Research Council of Canada (NSERC). Osmar Zaiane, a Canada CIFAR AI Chair, is also funded by the Canadian Institute for Advanced Research (CIFAR).

## References

[1] J. Mansky, The age-old problem of "fake news", 2018. URL: https://www.smithsonianmag.com/history/age-old-problem-fake-news-180968945/, accessed on 2021-11-24.
[2] Methodology, 2021. URL: https://mediabiasfactcheck.com/methodology/, accessed on 2021-11-24.
[3] International fact-checking network, 2021. URL: https://www.poynter.org/ifcn/, accessed on 2021-11-24.
[4] Verified signatories of the ifcn code of principles, 2021. URL: https://ifcncodeofprinciples.poynter.org/signatories, accessed on 2021-11-24.
[5] S. Xu, R. Huang, L. S. Sy, S. C. Glenn, D. S. Ryan, K. Morrissette, D. K. Shay, G. Vazquez-Benitez, J. M. Glanz, N. P. Klein, et al., Covid-19 vaccination and non–covid-19 mortality risk—seven integrated health care organizations, united states, december 14, 2020–july 31, 2021, Morbidity and Mortality Weekly Report 70 (2021) 1520.

[6] S. Mishra, S. Suryavardan, A. Bhaskar, P. Chopra, A. Reganti, P. Patwa, A. Das, T. Chakraborty, A. Sheth, A. Ekbal, C. Ahuja, Factify: A multi-modal fact verification dataset, in: Proceedings of the First Workshop on Multimodal Fact-Checking and Hate Speech Detection (DE-FACTIFY), 2022.

[7] D. Küçük, F. Can, Stance detection: A survey, ACM Computing Surveys 53 (2020) 1–37.

[8] A. Hanselowski, A. PVS, B. Schiller, F. Caspelherr, D. Chaudhuri, C. M. Meyer, I. Gurevych, A retrospective analysis of the fake news challenge stance-detection task, in: Proceed-ings of the 27th International Conference on Computational Linguistics, Association for Computational Linguistics, Santa Fe, New Mexico, USA, 2018, pp. 1859–1874. URL: https://aclanthology.org/C18-1158.

[9] B. Riedel, I. Augenstein, G. P. Spithourakis, S. Riedel, A simple but tough-to-beat baseline for the fake news challenge stance detection task, arXiv preprint arXiv:1707.03264 (2017).

[10] N. Reimers, I. Gurevych, Sentence-bert: Sentence embeddings using siamese bert-networks, in: Proceedings of the 2019 Conference on Empirical Methods in Natural Language Processing, Association for Computational Linguistics, 2019. URL: https://arxiv.org/abs/1908.10084.

[11] F. Chollet, Xception: Deep learning with depthwise separable convolutions, in: Proceedings of the IEEE conference on computer vision and pattern recognition, 2017, pp. 1251–1258.

[12] F. Chollet, et al., Keras, https://keras.io, 2015.

[13] G. Hinton, O. Vinyals, J. Dean, Distilling the knowledge in a neural network, arXiv preprint arXiv:1503.02531 (2015).

[14] Y. Pan, D. Sibley, S. Baird, Fake News Challenge - Team SOLAT IN THE SWEN, https://github.com/Cisco-Talos/fnc-1, 2018.

[15] T. Chen, C. Guestrin, XGBoost: A scalable tree boosting system, in: Proceedings of the 22nd ACM SIGKDD International Conference on Knowledge Discovery and Data Mining, KDD '16, ACM, New York, NY, USA, 2016, pp. 785–794. URL: http://doi.acm.org/10.1145/ 2939672.2939785. doi:10.1145/2939672.2939785.

[16] A. Hanselowski, A. PVS, B. Schiller, F. Caspelherr, athene_system, https://github.com/ hanselowski/athene_system, 2018.

[17] S. Özcan, santiagonasar, Rusty, Rushat, L. Lopez, Piotti, Preprocessor, https://github.com/ s/preprocessor, 2020.